\documentclass[floatfix,reprint,aip]{revtex4-1}
\usepackage{graphicx}% Include figure files
\usepackage{dcolumn}% Align table columns on decimal point
\usepackage{bm}% bold math
\usepackage{color,soul}
\usepackage{easyReview}
\usepackage[flushleft]{threeparttable}
\usepackage[utf8]{inputenc}
\usepackage[T1]{fontenc}
\usepackage{mathptmx}
\usepackage{etoolbox}
\usepackage{multirow}
\usepackage{hyperref}
\usepackage{mciteplus}
\usepackage{footnote}
\usepackage{tabularx}
\usepackage{color}
\usepackage{graphicx}
\usepackage{float}
\usepackage{amsmath}
\usepackage{amssymb}
\usepackage{amsfonts}
\makeatletter
\def\@email#1#2{%
 \endgroup
 \patchcmd{\titleblock@produce}
  {\frontmatter@RRAPformat}
  {\frontmatter@RRAPformat{\produce@RRAP{*#1\href{mailto:#2}{#2}}}\frontmatter@RRAPformat}
  {}{}
}%
\force@deferlist@sw{}{; try class option [floatfix]}
\makeatother
\begin{document}

\preprint{AIP/123-QED}

\title{Nanoextraction based on surface nanodroplets for chemical preconcentration and determination}
% Force line breaks with \\
\author{Hongyan Wu}
\affiliation{Department of Chemical and Materials Engineering, University of Alberta, Alberta T6G 1H9, Canada}
\author{Chiranjeevi Kanike}%
\affiliation{Department of Chemical and Materials Engineering, University of Alberta, Alberta T6G 1H9, Canada}
\affiliation{Department of Chemical Engineering, Indian Institute of Technology Kharagpur, Kharagpur, West Bengal 721302, India}
\author{Arnab Atta}%
\affiliation{Department of Chemical Engineering, Indian Institute of Technology Kharagpur, Kharagpur, West Bengal 721302, India}
\author{Xuehua Zhang}
 \homepage{https://sites.google.com/view/soft-matter-interfaces/home}
 \email{xuehua.zhang@ualberta.ca}
 \affiliation{Department of Chemical and Materials Engineering, University of Alberta, Alberta T6G 1H9, Canada}
\affiliation{Physics of Fluids Group, Max Planck Center Twente for Complex Fluid Dynamics, JM Burgers Center for Fluid Dynamics, Mesa+, Department of Science and Technology,University of Twente, Enschede 7522 NB, The Netherlands
}%

\date{\today}

\begin{abstract}

Liquid-liquid extraction based on surface nanodroplets, namely nanoextraction, can continuously extract and enrich target analytes from the ﬂow of a sample solution. This sample preconcentration technique is easy to operate in a continuous flow system with a low consumption of organic solvent and a high enrichment factor. In this review, the evolution from single drop microextraction to advanced nanoextraction will be briefly introduced. Also, the formation principle and key features of surface nanodroplets will be summarized. Further, the major findings of nanoextraction combined with in\textendash droplet chemistry towards sensitive and quantitative detection will be discussed. Finally, we will give our perspectives for the future trend of nanoextraction. 

\end{abstract}

\maketitle

\section{Introduction}

Chemical analysis at trace levels is crucial in various fields such as food safety screening, \cite{wang2021emerging,choi2019emerging} environmental pollutant monitoring, \cite{wong2021nanozymes,zhu2018fluorescent} clinical forensics, \cite{sauvage2006screening,steuer2019metabolomic} and biological contaminants sensing. \cite{rasheed2019environmentally,liu2018hundreds} A preliminary step of analyte extraction and enrichment prior to detection contributes to improve the limit of detection (LoD) of analytical tools.\cite{aguirre2015dispersive,ramos2012critical} Due to the large active surface area, microscopic droplets are of great interest for preconcentration of target analytes across the droplet-liquid interface at extremely low concentrations.\cite{henschke1999mass,anthemidis2009development, feng2018droplet, zhang2022biphasic} The enhanced preconcentration depends on the partition of the compound between droplet liquid and surrounding sample liquid.

Recently, dispersive liquid-liquid microextraction (DLLME) has received great attention because of its rapid and efficient extraction. \cite{rezaee2006determination,galuch2019determination,lemos2022syringe,gallo2021dispersive,grau2022use,carbonell2021natural,shojaei2021application,ji2021hydrophobic,wang2021strategies} DLLME is a spontaneous emulsification technique based on a ternary mixture containing dispersive solvent, extractant, and aqueous sample solution. \cite{vitale2003liquid,rezaee2009dispersive} The formation of extractant microdroplets increases the interfacial mass transfer of the analytes, leading to increased extraction efficiency. The microdroplets are then centrifuged and collected from the bulk for subsequent analysis. The main disadvantages of DLLME are the requirement of two individual steps for extraction and equipment-assisted sample separation, as well as the consumption of relatively large amount of disperser solvents. 

Surface nanodroplets on immersed substrates provide an alternative platform for efficient extraction. The most-used method to induce surface nanodroplets is the solvent exchange. In this process, surface nanodroplet nucleation and subsequent growth occur due to the droplet liquid transient oversaturation when a good solvent of the droplet liquid is displaced by a poor solvent. \cite{zhang2015formation} Surface nanodroplet exhibits huge surface-to-volume ratio and excellent long-term stability against evaporation or dissolution. \cite{li2020speeding, an2015wetting} These features enable surface nanodrops to continuously extract and enrich trace amounts of solutes from the ﬂow of an aqueous solution. The extraction efficiency of surface nanodroplets produced on a planar or a curved surface was investigated by Yu et. al in 2016. \cite{yu2016large} The term ‘nanoextraction’ describing the liquid–liquid extraction based on surface nanodroplets was proposed by Li et al. in 2019. \cite{li2019functional}

Nanoextraction can be integrated with surface reactions for further application in chemical analysis. It has been demonstrated that many chemical reactions in nanodroplets are faster and more effective than their macroscopic counterparts in a bulk medium. \cite{zhang2016compartmentalized,piradashvili2016reactions} Reactive components could be introduced to a single droplet to impart multi\textendash functionalities under well-controlled mixing conditions without the influence of uncontrolled collision, coalescence or Ostwald ripening for droplet reaction in a bulk system. Nanoextraction coupled with in-situ surface reactions provides an active platform for synthesis of functionalized nanomaterials and combinative analysis.\cite{li2019functional,dabodiya2022sequential}

This review aims to summarize the latest progress and point out the current research trends of nanoextractions. Herein, we begin by briefly introducing the historical development of droplet-based extraction, followed by elaborating the formation principle and control parameter of surface nanodroplets. Then, we discuss the current research progress of nanoextraction, as well as reaction-assisted nanoextraction. Finally, we give our perspectives of what would be the future development of nanoextraction in terms of analytical practices. Prior review articles on surface nanodroplet mainly focus on its formation and dissolution dynamics. \cite{lohse2015surface,qian2019surface} To the best of our knowledge, there has been no review article discussing the nanoextraction based on surface nanodroplet. We hope this review could provide support for further development of nanoextraction in the trace chemical component analysis.

\begin{figure}[htbp]
\centering
 \includegraphics[height=8cm]{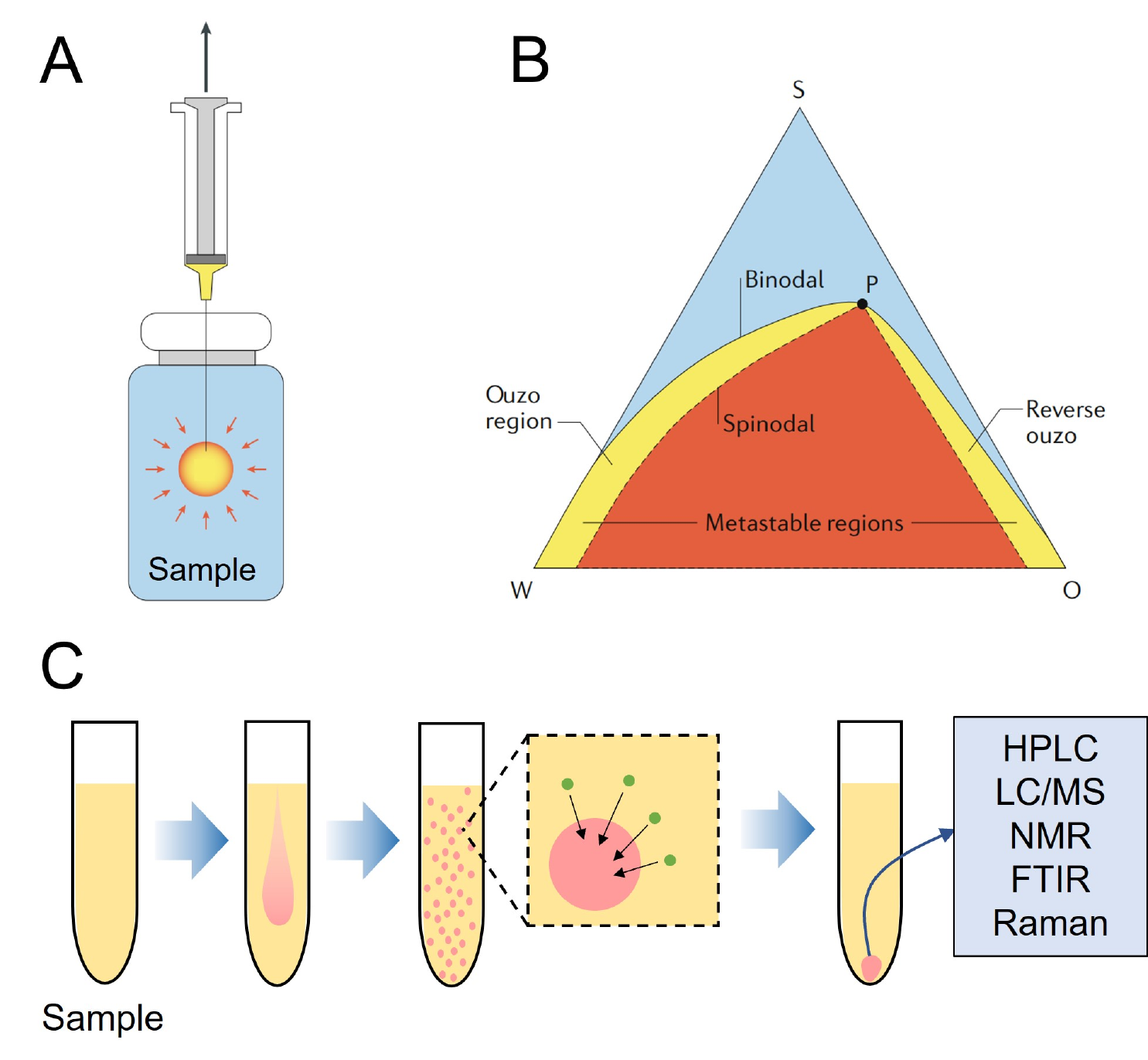}
  \caption{(A) Illustration of the classical single drop microextraction (SDME) process. (B) Schematic illustration of a ternary phase diagram of oil (O), water (W), and solute (S). Oil or water droplets can be formed in the ouzo region or reverse ouzo region, respectively.  (C) General Schematic of the classical DLLME procedure. Panel (A) and (B) taken from ref.\cite{lohse2020physicochemical} Copyright 2020 Springer Nature. Panel (C) reproduced with permission from ref.\cite{rezaee2010evolution} Copyright 2009 Elsevier B.V. }
\label{fgr:History}
\end{figure}

\section{Development of droplet-based extraction}

Single drop microextraction (SDME), as a solvent miniaturized extraction technique, was explored by Liu et al.\cite{liu1996analytical}  and Jeannot et al. \cite{jeannot1996solvent} in 1996. In a typical SDME process, an organic microdroplet ($\sim$1\textendash 10 $\mu$L) is suspended at the tip of a syringe needle which is surrounded by an aqueous sample solution (Fig. \ref{fgr:History}A). After extraction, the drop with enriched analytes is withdrawn and injected into an analytical instrument for detection and quantification. The extraction of target compounds from aqueous phase into an organic solvent drop is based on passive diffusion. \cite{yangcheng2006directly,sikanen2010implementation} The extraction efficiency is intrinsically determined by the analyte partition coefficient between the droplet liquid and water. Different approaches to perform SDME, such as  continuous-flow microextraction, \cite{liu2000continuous,wu2016dynamic} headspace SDME, \cite{vidal2005headspace,snow2010novel} drop-to-drop microextraction, \cite{wijethunga2011chip,shrivas2007rapid} and direct immersion-SDME, \cite{ruiz2014ternary,nunes2020direct} have been extensively explored for analytical applications.  Jeannot et al.\cite{jeannot2010single} summarized the historical development and pointed out advantage/disadvantages of various modes of SDME technique. A recent review from 2021 focused on the applications of SDME combined with multiple analytical tools (spectroscopy, chromatography, and mass spectrometry).\cite{kailasa2021applications} The popularity of SDME mainly lies in its cost-effectiveness and low consumption of organic solvent. Its drawbacks are the instability of the hanging droplet, narrow drop surface, and consequently slow diffusion kinetics and limited sensitivity.

DLLME, as a modified solvent microextraction technique, was introduced in 2006 by Rezaee et al. for the extraction of organic analytes from aqueous samples. \cite{rezaee2006determination} In this method, multiple extractant microdroplets are formed and stably dispersed in an aqueous sample comprising of the target analytes. The formation of small droplets is based on the spontaneous emulsification in a ternary system, which is well known as the “Ouzo effect”. \cite{tan2016evaporation} A typical ternary mixture consists of a good solvent (e.g., ethanol), a poor solvent (e.g., water), and a small ratio of extractant (e.g., oil). The three-phase diagram of a representative ternary system is shown in Fig. \ref{fgr:History}B. The nucleation and growth of the droplets spontaneously occur in the ouzo regime, which is surrounded by the binodal and spinodal curves. \cite{aubry2009nanoprecipitation} This emulsification without surfactant is kinetically stable over hours or days.\cite{prevost2021spontaneous} The dramatically increased surface area leads to the enhanced diffusion kinetics and high recovery efficiency.

The typical procedure of DLLME is demonstrated in Fig. \ref{fgr:History}C, microdroplets of extractant are spontaneously formed when a mixture containing an extracting solvent and a disperse solvent is rapidly injected into an aqueous sample solution. The partition equilibrium at the droplet interface is reached in a few seconds owing to the large surface area. Consequently, the extraction is almost independent of time.\cite{assadi2010determination} The cloudy emulsion is then centrifuged to collect the droplets with concentrated target compound for analysis. Over the years, DLLME technique has been developed from its basic approach into many other advanced configurations. Recent revolutions of DLLME have been made regarding to the selection of extracting and disperse solvent,\cite{deng2019hexafluoroisopropanol,el2019deep} combination with other extraction techniques,\cite{shamsipur2015extraction,rai2016comparative} association with derivatization reaction,\cite{sajid2018dispersive,pinto2018quantitative} mechanical agitation-assisted emulsification, \cite{mansour2017solidification,de2015new} etc. The review devoted specifically to DLLME was published by Rezaee et al. \cite{rezaee2010evolution} in 2010 comprehensively summarizing the early development of DLLME. Sajid et al. \cite{sajid2022dispersive} reviewed latest advancements of DLLME with respect to its evolved design of devices, green aspects, and application extensions. Compared to SDME, DLLME greatly improves the enrichment efficiency. Some limitations still exist, such as the consumption of toxic extraction solvents (e.g., carbon tetrachloride, chlorobenzene, and cyclohexane).\cite{rezaee2010evolution}

\begin{figure}[htbp]
\centering
 \includegraphics[height=9.5cm]{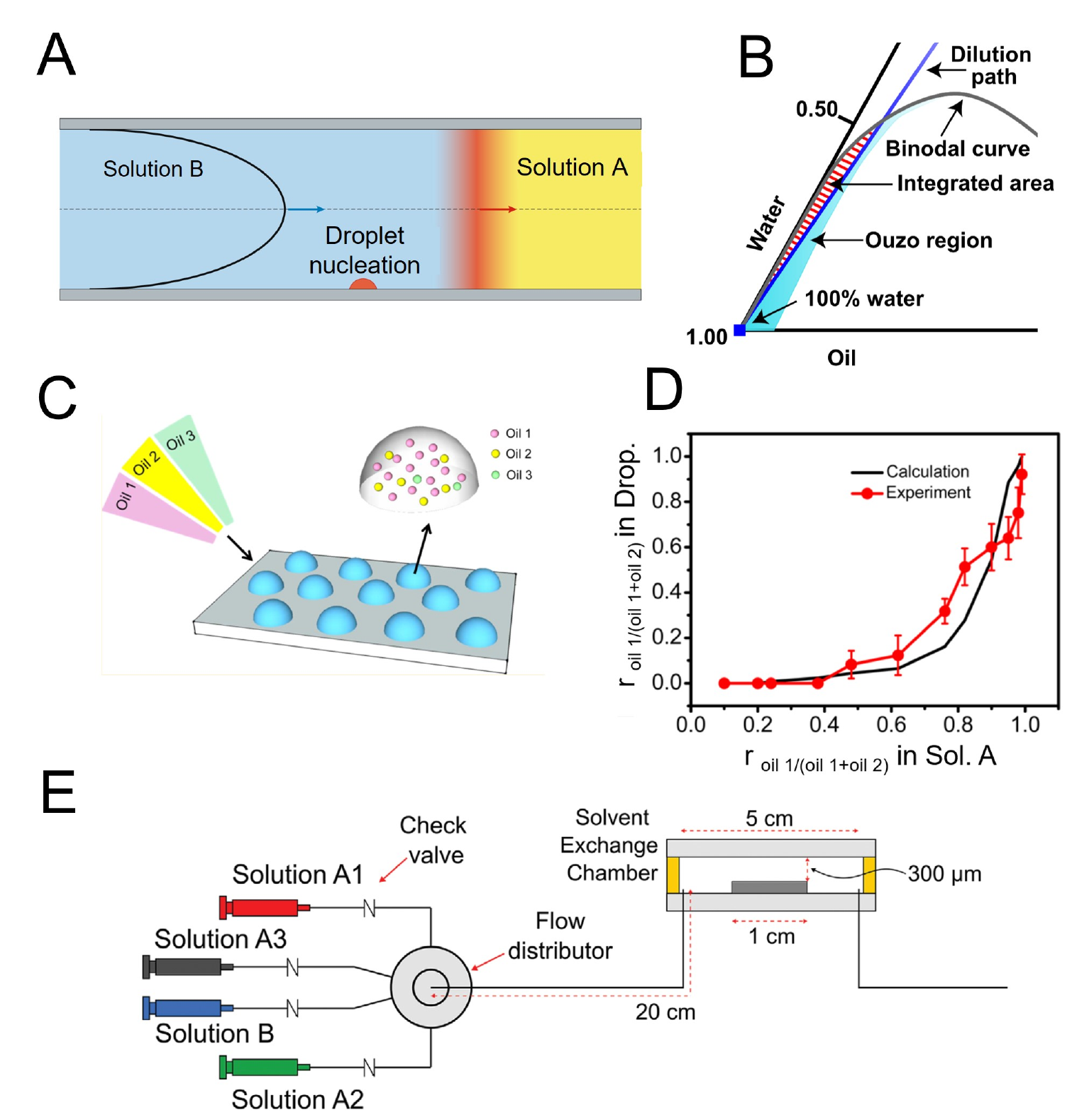}
  \caption{(A) Schematic showing the typical solvent exchange process. (B) Sketch showing the solution composition during the formation of oil nanodroplets in the solvent exchange. The dilution path represents the solution composition with a oil-to-ethanol ratio same as in solution A. The overall oil supply (oversaturation level) is reflected by the shaded area. (C) Sketch showing the multicomponent surface nanodroplets comprising three different oils. (D) Plots of oil composition in the binary droplets compared to their initial ratio in solution A. (E) Schematic illustration of the continuous mixing setup for the formation of multicomponent surface nanodroplets. The flow distributor is used to mix different solution $A_i$ (i = 1, 2, 3) and  introduce the mixture to a solvent exchange chamber. Panel (A) taken from ref.\cite{lohse2020physicochemical} Copyright 2020 Springer Nature. Panel (B) reprinted with the permission from ref. \cite{lu2015solvent} Copyright 2015 American Chemical Society. Panel (C) and (D) reproduced with the permission from ref.\cite{li2018formation} Copyright 2018 American Chemical Society. Scheme (E) taken from ref. \cite{you2021tuning} Copyright 2021 John Wiley \& Sons, Inc. }
\label{fgr:Formation}
\end{figure}

\section{Formation of surface nanodroplets}

The formation principle of surface nanodroplets is also based upon the “ouzo effect”.\cite{zhang2015formation} The standard protocol of the solvent exchange to form surface nanodroplets is shown in Fig. \ref{fgr:Formation}A. Different from the generation of dispersive microdroplets in a DLLME process, surface nanodroplets are induced by an oversaturation pulse at the interacting front of the ouzo solution (solution A) and the poor solvent (solution B) in a narrow chamber.\cite{lohse2020physicochemical} As shown in Fig. \ref{fgr:Formation}B, the oversaturation level is demonstrated by the integrated area between the binodal curve and the dilution path through the ouzo region. The nanodroplets pinned on the substrate are 5\textendash 500 nm in height and 0.1\textendash 10 $\mu$m in lateral diameter. \cite{li2019controlled}

The surface properties (e.g., hydrophobicity, patterns, etc.) of the substrate will affect the droplet formation, growth, and distribution.\cite{bao2015highly,lohse2015surface} The wettability of the substrate needs to be compatible with the droplet liquid. Oil droplet can be formed on the hydrophobic substrate, while water droplet can be formed on the hydrophilic substrate. The prepatterned microdomain on the substrate defines the spatial arrangement and lateral dimension of the surface droplets.\cite{bao2015highly} The growth of the nucleated droplets follows the constant contact angle mode when the lateral diameter is smaller than that of the domain. After the droplets reach the domain circumference, growth proceeds in constant contact radius mode until the solvent exchange is completed. Finally, highly ordered array of surface nanodroplets with well\textendash defined sizes are generated on the patterned substrate. For those unpatterned substrates, surface nanodroplets are sparsely and 
heterogeneously distributed over
the whole area. \cite{zhang2012transient} 

A variety of liquids, such as water\cite{lu2016influence, wei2020integrated}, alkanes, \cite{lu2015solvent, lu2016influence} alcohols,\cite{dyett2018coalescence,dyett2018growth} fatty acids,\cite{lu2017universal} and other oils,\cite{tan2016evaporation} can be used to produce surface nanodroplets based on an appropriate ternary system. The formation of nanodroplets are tunable in the solvent exchange. More specifically, the solvent type and solution composition influence the oversaturation in the three-phase diagram and consequently the number density and size of the nanodroplets.\cite{lu2015solvent, lu2016influence} The droplet diameter increases almost linearly with the square root of the shaded area in Fig. \ref{fgr:Formation}B. For a given ternary system, the droplet size can be controlled by the flow rate and the chamber geometry.\cite{zhang2015formation,dyett2018growth} Qian et al. \cite{qian2019surface} systematically summarized the effects of different factors on the formation of nanodroplets. The flexibility of droplet liquid, tunable droplet distribution and size, and long-term stability make surface nanodroplets good candidates for extraction. 

Multicomponent surface droplets comprising more than one compound can also be produced by solvent exchange (Fig. \ref{fgr:Formation}C).\cite{li2018formation} In this process, different droplet liquids are mixed at a certain ratio in solution A prior to the solvent exchange. As is indicated in Fig. \ref{fgr:Formation}D, the oil 1 content (oil 1/ (oil 1 + oil 2)) in the binary droplets varies from 0 to 1 by changing the initial ratio of these two liquids in the solution. Nevertheless, the ratio of these two oils in the droplets is not directly formulated by their composition in the original solution but is governed by the oversaturation ratios of each component in the three\textendash phase diagram.\cite{li2018formation} Following the same principle, ternary droplets with desired ratio can be generated by controlling the oversaturation of each oil. 

You et al. \cite{you2021tuning} developed a continuous flow-in setup to further simplify the formation and formulation of multicomponent droplets. As is shown in Fig. \ref{fgr:Formation}E, a flow distributor is employed as a passive mixer to introduce more than one streams of solution A during the solvent exchange. The composition of the produced surface nanodroplets can be easily tuned by adjusting the flow rate ratios of each stream of the solution A with a high reproducibility. These controllable multicomponent surface droplets are of great interest in the compartmentalized chemical reactions and microanalytics.\cite{zheng2021microfluidic,lohse2020physicochemical}

\begin{figure*}[htbp]
\centering
 \includegraphics[height=14cm]{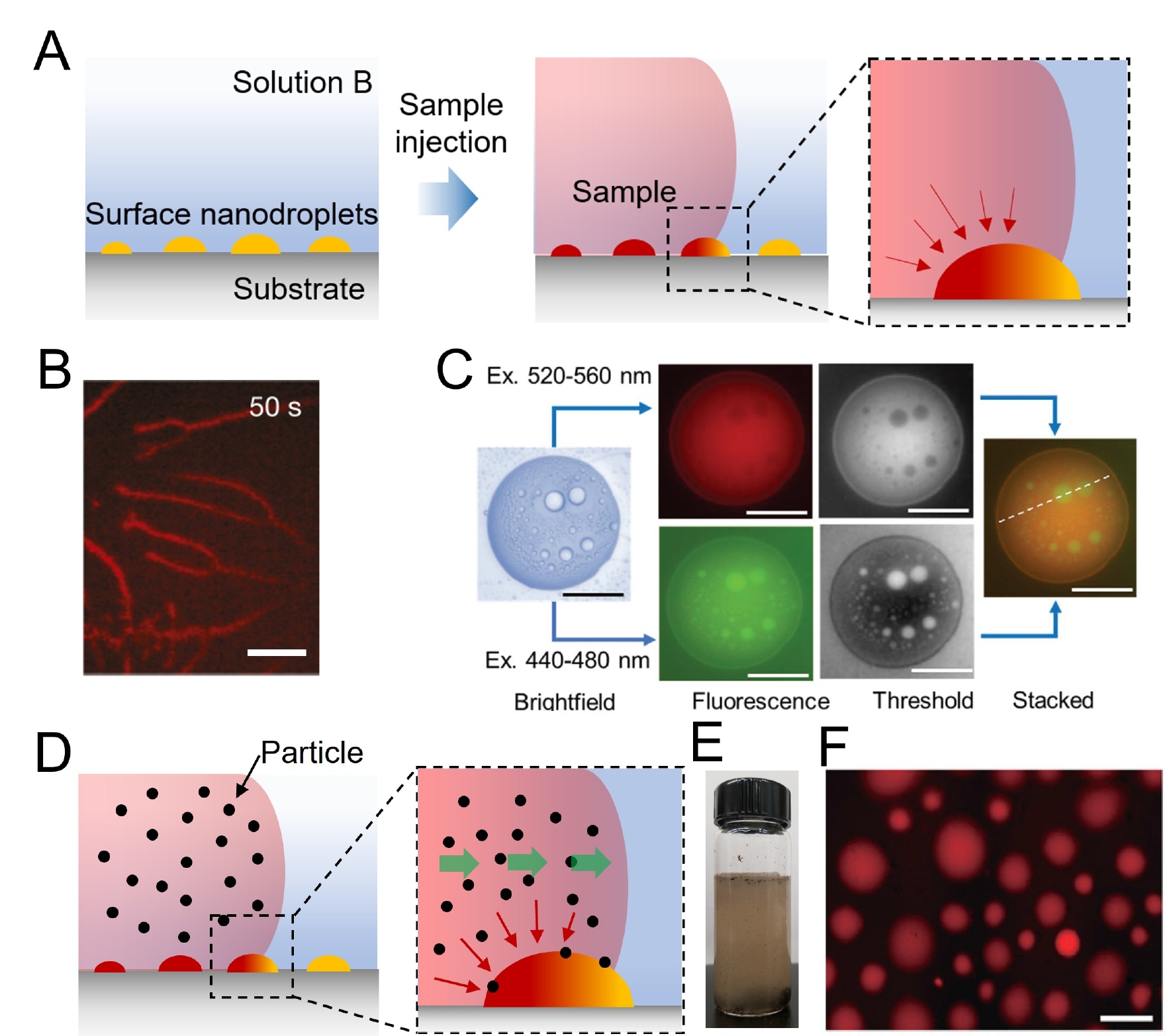}
  \caption{ (A) Schematic illustration showing the extraction of analytes from the sample flow into the surface nanodroplets.  (B) Fluorescent image showing a red dye in water (10 nM) is extracted into the surface droplet branches over time. Scale bar: 200 $\mu$m. (C) Fluorescent images showing selective extraction of biphasic surface droplets from a mixture of R6G and fluorescein under different ﬂuorescence excitations. Scale bar: 20 $\mu$m. (D) Sketch showing the extraction of analytes from the dense suspension into the surface nanodroplets. (E) Image of the oil sand wastewater consisting of bitumen, solid particles, and hydrocarbons. (F) Fluorescent image showing the surface nanodroplets after extracting 10$^{-6}$ M Nile red dye from the oil sand wastewater. ($\sim$ 30 wt\% solid content). Scale bar: 100 $\mu$m. Scheme (A) reproduced with the permission from ref. \cite{yu2016large} Copyright 2016 American Chemical Society. Panel (B) taken from ref. \cite{lu2017universal} Copyright 2022 National Academy of Science. Panel (C) taken from ref.\cite{li2020encapsulated} Copyright 2020 John Wiley \& Sons, Inc. Panel (D)\textendash(F) reproduced with the permission of ref. \cite{you2021surface} Copyright 2021 The Royal Society of Chemistry. }
\label{fgr:Extraction-1}
\end{figure*}

\begin{figure*}[htbp]
\centering
 \includegraphics[height=10cm]{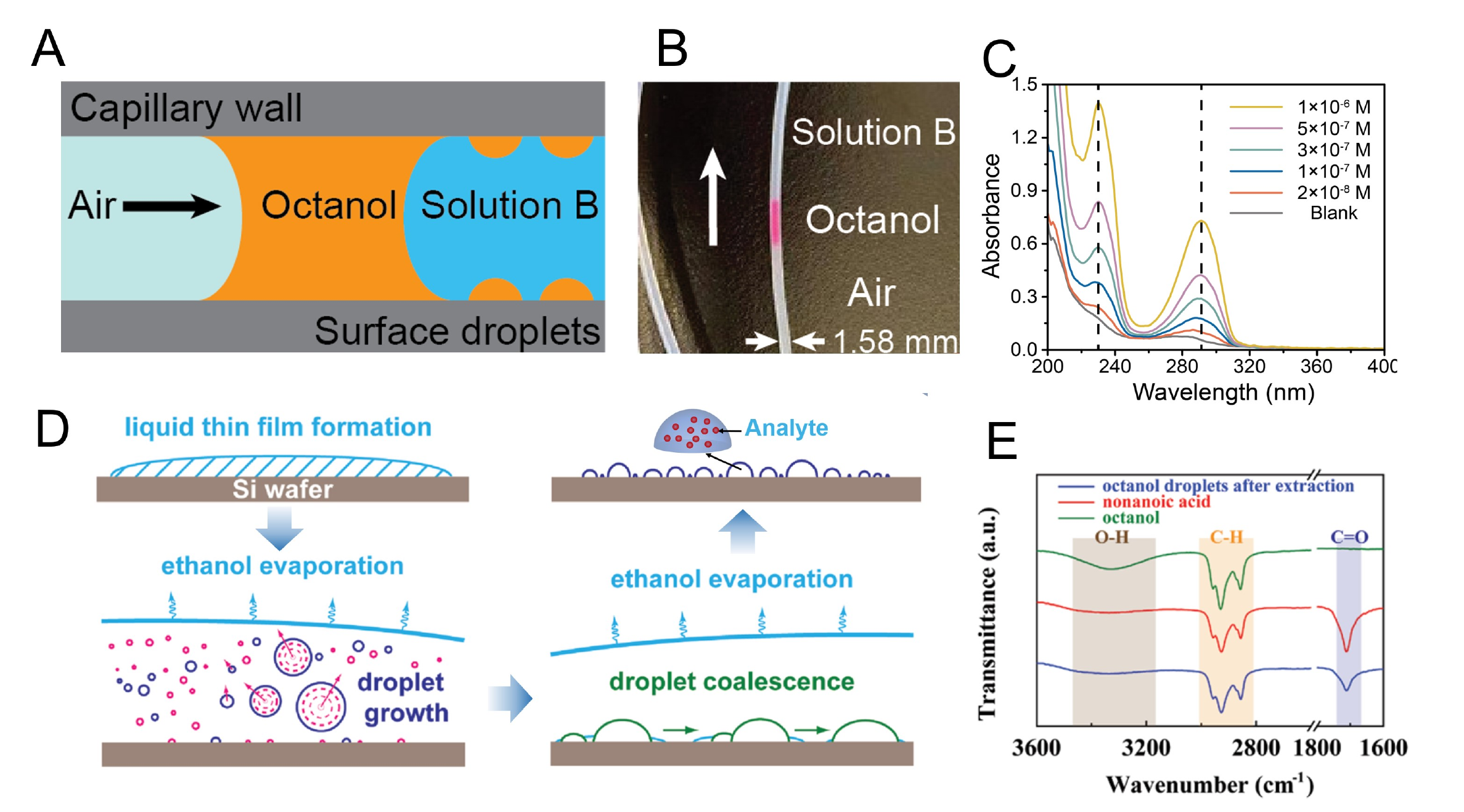}
  \caption{ (A) Schematic showing the collection of surface nanodroplets by the capillary force between air and water. (B) Image showing the collected octanol in the capillary tube. (C) UV-Vis absorbance spectra of collected octanol droplets after the extraction from aqueous sample flow with chlorpyrifos concentrations
 from $2\times10^{-8}$ M to 10$^{-6}$ M. (D) Schemetic illustraion showing the fomarion process of the surface nanodroplets in an evaporating liquid film (E) FTIR spectra of octanol droplets after extraction of 10$^{-3}$ {M} nonanoic acid, octanol, and nonanoic acid. Panel (A), (B), and (C) taken from ref. \cite{li2022surface} Copyright 2022 American Chemical Society. Panel (D) and (E) reproduced with the permission from ref. \cite{qian2020one} Copyright 2019 John Wiley \& Sons, Inc.}
\label{fgr:Extraction-2}
\end{figure*}

\section{Nanoextraction based on surface nanodroplets}

\citet{yu2016large} illustrated the feasibility of surface nanodroplets for nanoextraction. Highly ordered surface droplets were generated on a vertical prepatterned substrate and used to extract the model compound rhodamine 6G (R6G) from the flow (Fig. \ref{fgr:Extraction-1}A). The enrichment efficiency was quantified by the fluorescent intensity ratio between the oil droplets and the surrounding media. The results showed that the concentrations of R6G in oil droplets were increased by a factor of 3, indicating the significant enrichment of the hydrophobic solute from the highly diluted aqueous solution. \citet{lu2017universal} studied the droplet branches confined in quasi-2D channel for nanoextraction. As is illustrated in Fig. \ref{fgr:Extraction-1}B, the fluorescent intensity of nanodroplet branches gradually increased over time, demonstrating the red dye in water is continuously extracted and accumulated into the oil nanodroplet branches. 

Li et al. \cite{li2020encapsulated} produced biphasic surface droplets within water nanodroplets enclosed in oil microdroplets. This biphasic feature of the droplet unit allows selective concentration of both hydrophobic and hydrophilic analytes from a mixture flow. Fig. \ref{fgr:Extraction-1}C demonstrates the selective ﬂuorescence enhancement of the host and encapsulated droplets. An aqueous mixture containing hydrophobic R6G and hydrophilic ﬂuorescein was introduced to the host droplet for extraction. Those two analytes can be separately excited at different range of wavelength. R6G is concentrated into the host droplet while ﬂuorescein is enriched into the encapsulated nanodroplets. The compartmentalized nature of the biphasic droplets may contribute to broader applications in sensing and diagnostic fields.

You et al. \cite{you2021surface} developed the nanoextraction in a capillary tube. Due to the narrow diameter of the capillary, reliable extraction is achieved with a sample volume as low as 50 $\mu$L. This method enables extraction of trace compounds from the slurry with highly concentrated solid particles. As is demonstrated in Fig. \ref{fgr:Extraction-1}D. The target compound in a suspension sample is steadily extracted into the surface nanodroplets. Although part of solid particles is adsorbed and aggregated at the droplet-water interface, their influence on the extraction and in-situ detection would be minimal. The LoD of a model compound (Nile red) for aqueous samples and slurry samples was found to be 10 nM and 1 nM, respectively. Further, the successful extraction and detection of analytes from a oil sand wastewater ($\sim$ 30 wt\% solid content) indicates the potential of this technique in the application of environmental monitoring (Fig. \ref{fgr:Extraction-1}E,F).

In addition to in-situ analysis, the surface nanodroplets with extracted analytes in a 3-meter hydrophobic capillary tube could be collected (total volume $\geq$ 2 $\mu$L) for ex-situ detection.\cite{li2022surface} The collection of surface nanodroplets is governed by the capillary force between water and air (Fig. \ref{fgr:Extraction-2}A,B). The collected octanol was then analyzed by a UV-vis spectrometry (Fig. \ref{fgr:Extraction-2}C). The LoD was found to be $\sim$ $2\times10^{-9}$ M for two representative micropollutants, triclosan and chlorpyrifos. A linear range above 2 × 10$^{-7}$ M was achieved for quantification. This nanoextraction method was also proved to be compatible with GC-MS and fluorescence microscopy.

The formation of surface nanodroplet is based not only on the solvent exchange method but also on an evaporating ternary liquid microfilm. Qian et al.\cite{qian2020one} investigated the nanoextraction efficiency in an evaporating thin liquid film on a spinning substrate. In brief, in a ternary system consisting of a target analyte in water, an extractant oil and a co-solvent ethanol, nanodroplets of oil formed spontaneously in the film within rapid evaporation of ethanol (Fig. \ref{fgr:Extraction-2}D). In this case, an even less amount of analyte solution (5 $\mu$L) is required, and the entire process could be completed in 10 s. The LoD of the model compound Nile Blue was found to be 10$^{-12}$ M by in-situ fluorescence detection. Besides, droplets with extracted nonanoic acid enabled in-situ chemical identification and quantification by a FTIR microscope. (Fig. \ref{fgr:Extraction-2}E). The successful online and offline detection of surface nanodroplets by various analytical tools indicates that nanoextraction can be a general approach for sample preconcentration.

\section{Nanoextraction coupled with in-droplet reaction}

\begin{table*}[htbp]
  \begin{threeparttable}
    \caption{ Level of enhancement $p/p^*$ of Long-Chain Alkyl Acid. Reproduced with the permission of ref.\cite{wei2022interfacial} Copyright 2022 American Chemical Society. }
    \begin{tabular*}{1\textwidth}{@{\extracolsep{\fill}}cccccc}
      \hline
Acid                  & 0.5 mM & 1 mM & 5 mM & 50 mM & 100 mM \\
\hline
Acetic acid           & 2.25   &      & 1.01 & 1     &        \\
N-butyric acid        &        & 11.4 & 3.4  & 1     &        \\
Gallic acid           &        & 4.2  & 1.1  & 1     &        \\
4-hydroxybenzoic acid &        &      & 7.3  & 1.5   & 1      \\
Benzoic acid          &        &      & 6.2  & 1.2   & 1 \\    
 \hline
\end{tabular*}
  
  \end{threeparttable}
\end{table*}

\begin{figure*}[htbp]
\centering
 \includegraphics[height=12cm]{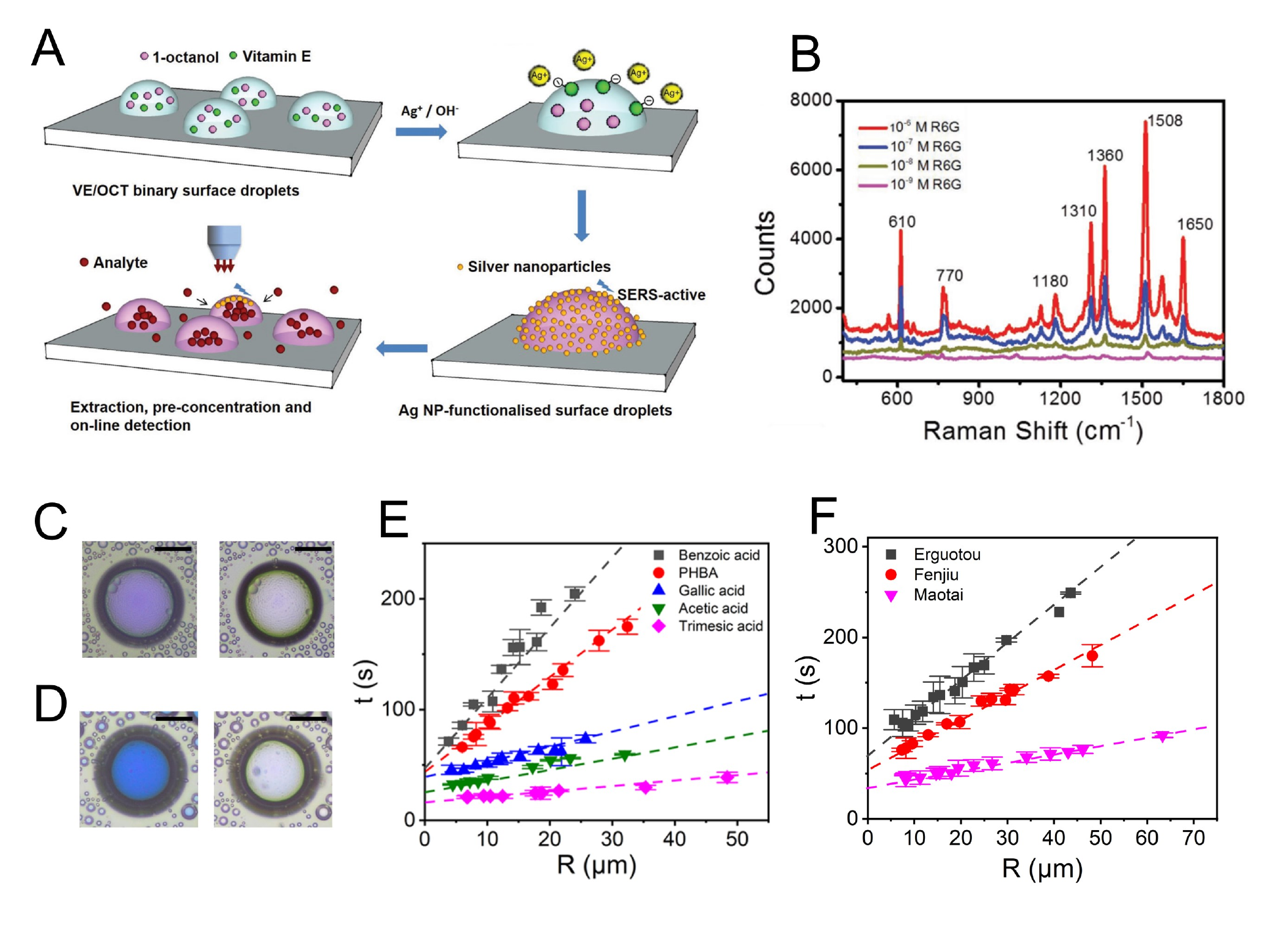}
  \caption{ (A) Schematics showing the fabrication of AgNPs, the enrichment of analytes, and the in-situ SERS detection based on surface droplets. (B) Raman spectra of surface nanodroplets after the extraction from aqueous sample flow with R6G concentrations from 10$^{-9}$ M to 10$^{-6}$ M. (C) Optical images of a droplet containing bromocresol purple before and after interact with $1.7\times 10^{-4}$ M acetic acid. (D) Optical images of a droplet containing bromocresol green before and after interact with $1.0\times 10^{-4}$ M acetic acid. Length of the scale bar: 25 $\mu$m. (E) Droplet discoloration time versus the reciprocal of droplet size for different types of acid. The acid concentration is 5 mM. (F) Detection of different Chinese spirits by surface droplets. Panel (A) and (B) taken from ref. \cite{li2019functional} Copyright  2019 John Wiley \& Sons, Inc. Panel (C)\textendash(F) taken from ref.\cite{wei2020integrated} Copyright 2022 American Chemical Society.}
\label{fgr:Reaction}
\end{figure*}

Nanoextraction can be combined with surface reactions for online chemical analysis. \cite{li2019functional} As shown in Fig. \ref{fgr:Reaction}A, binary droplets comprising reactive component (Vitamin E) and nonreactive component (1-octanol, OCT) were produced by solvent exchange process. AgNO$_3$ aqueous solution were then injected to react with Vitamin E in the droplets and generate silver nanoparticles (AgNPs). OCT in droplets acts as a preconcentration agent to extract analyte from a continuous flow. In situ sensitive surface-enhanced Raman spectroscopy (SERS) detection was achieved based on the AgNPs-functionalized droplets (Fig. \ref{fgr:Reaction}B). LoD reaches to 8 $\times$ 10$^{-11}$ M for methylene blue, 3 $\times$ 10$^{-9}$ M for malachite green, and 10$^{-10}$ M for R6G, respectively. Besides, the quantitative and reproductive detection of R6G was achieved within a large linear range from 10$^{-9}$ to 10$^{-6}$ M.

\citet{wei2020integrated} integrated nanoextraction and colorimetric reactions for acid detection by droplet decoloration time. The formed aqueous droplets contain two halochromic compounds, bromocresol green and bromocresol purple. The acid dissolved in an oil flow was extracted into the water droplets. The reaction of the extracted acid with the halochromic compound leads to the pH change and subsequent decoloration of the droplets over a certain time. The decoloration can be simply identified by an optical microscope (Fig. \ref{fgr:Reaction}C,D). It is clear in Fig. \ref{fgr:Reaction}E that the time for droplet decoloration is dependent on the type of acid. This chemical specificity allows the identification of different acid mixture. The authors successfully applied this drop-based analysis method for distinguishing counterfeit alcoholic spirits by comparing their discoloration time. For example, the decoloration time of three Chinese famous spirits Fenjiu, Maotai, and Erguotou is distinguishable due to their characteristic acid profile (Fig. \ref{fgr:Reaction}F). 

In a follow-up work performed by \citet{wei2022interfacial}, the authors experimentally and theoretically studied the mechanism behind the enhanced extraction by surface nanodroplets. The droplet
decoloration process is associated with three steps: 1) interfacial partition, 2) dissociation, and 3) colorimetric reaction. Step 1 is recognized as the rate-limiting step. Further, the concentration change rate of acid in droplets from those three steps can be quantified by: 
\begin{equation}
\frac{dC_w(t)}{dt}=\frac{C}{R}\beta(\frac{C_o(t)}{p}-C_w(t))+rC_w(t)
\label{e1}
\end{equation}
The first term on the right represents interfacial transfer of acid (step 1). The second term represents reaction of acid in the droplets (step 2 and 3). From eqn (1), the shift from actual partition coefficient ($p$) to apparent partition coefficient ($p^*$) of acid is observed to correspond to more acid molecules extracted into the droplets. As is indicated in Table 1, The partition of acid in the reacting droplets can be shifted up to 11 times of that in the bulk by interfacial colorimetric reactions.

\section{Conclusions and outlook}
Advances have been made towards efficient liquid-liquid nanoextraction based on surface nanodroplets. The formation of nanodroplets can be achieved in a fluid chamber or a capillary tube by solvent exchange, or in an evaporating thin liquid film. Target analytes can be continuously extracted from a sample ﬂow into the pinned surface droplets. The extraction is almost not limited by the sample volume or affected by the solid particles. After extraction, nanodroplets with concentrated analytes can be directly used for online analysis (fluorescent microscope, FTIR microscope, and Raman spectroscopy) or be collected for offline analysis (GC-MS and Uv-vis spectroscopy). The stable nanodroplet with large surface-to-volume ratio is a general platform for interfacial reaction, which enhances the shift of partition and enables synthesis of functional materials for SERS detection. 

Nanoextraction has emerged as a powerful and reliable sample-preconcentration approach due to its high extraction efficiency, simplicity of operation, low cost and high environment benignity. This technique is easily accessible to most of researchers and laboratories. Further development of nanoextraction may be expanded to complex matrixes, such as biological samples (e.g., cells and tissues). The organic solvents in the ternary system are used for the formation of surface droplets. After droplet formation, the surrounding liquid can be aqueous solution. The biological samples may be brought into contact via their aqueous solution. The biocompatibility issues would be minimal in the nanoextraction process. 

\begin{acknowledgments}
We are grateful for our close collaborators for their valuable findings. We acknowledge funding support from the Natural Sciences and Engineering Research Council of Canada (NSERC)\textendash Discovery project, and Alliance Grant Alberta Innovates\textendash Advanced program. X.H.Z. acknowledges support from the Canada Research Chairs Program. H.Y.W acknowledges support from China Scholarship Council (No. 202106450020).
\end{acknowledgments}

\section*{Nomenclature}
\noindent $C_w(t)$: Acid concentration change rate in water droplet;\\
$C_o(t)$: Acid concentration change rate in oil flow;\\
$\beta$: Mass transfer coefficient;\\
$p$: Actual distribution coefficient;\\
$p^*$: Apparent distribution coefficient;\\
$r$: Total reaction rate in water droplet;\\
$R$: Lateral radius of acid surface droplets.

\section*{AUTHOR DECLARATIONS}
\subsection*{Conflict of Interest}
The authors have no conflicts to disclose. 
\section*{Data Availability}
Data sharing is not applicable to this article as no new data were created or analyzed in this study.

\section*{REFERENCES}
\nocite{*}
\bibliography{ref}

\end{document}